\documentclass[proceedings,conferenceproceedings,accept,moreauthors,pdftex,10pt,a4paper]{Definitions/mdpi} 

\firstpage{1} 
\pubvolume{xx}
\issuenum{1}
\articlenumber{5}
\pubyear{2018}
\copyrightyear{2018}
\history{Received: date; Accepted: date; Published: date}

\Title{Centrality dependence of deuteron production in Pb+Pb collisions at 2.76 TeV via hydrodynamics and hadronic afterburner$^{\dagger}$}

\Author{Dmytro~Oliinychenko $^{1}$\orcidA{}, Long-Gang~Pang$^{1,2}$\orcidB{}, Hannah~Elfner$^{3,4, 5}$\orcidC{} and Volker Koch$^{1}$\orcidD{}*}

\AuthorNames{Dmytro Oliinychenko, Long-Gang Pang, Hannah Elfner and Volker Koch}

\address{%
$^{1}$ \quad Lawrence Berkeley National Laboratory, 1 Cyclotron Rd, Berkeley, CA 94720, US\\
$^{2}$ \quad Physics Department, University of California, Berkeley, CA 94720, USA\\
$^{3}$ \quad GSI Helmholtzzentrum für Schwerionenforschung, Planckstr. 1, 64291 Darmstadt, Germany\\
$^{4}$ \quad Institute for Theoretical Physics, Goethe University, Max-von-Laue-Strasse 1, 60438 Frankfurt am Main, Germany\\
$^{5}$ \quad Frankfurt Institute for Advanced Studies, Ruth-Moufang-Strasse 1, 60438 Frankfurt am Main, Germany
}

\corres{Correspondence: doliinychenko@lbl.gov}

\firstnote{Presented at Hot Quarks 2018 - Workshop for young scientists on the physics of ultrarelativistic nucleus-nucleus collisions, Texel, The Netherlands, September 7-14 2018}

\abstract{The deuteron binding energy is only 2.2 MeV. At the same time, its yield in Pb+Pb collisions at $\sqrt{s_{NN}} = $2.76 TeV corresponds to a thermal yield at the temperature around 155 MeV, which is too hot to keep deuterons bound. This puzzle is not completely resolved yet. In general, the mechanism of light nuclei production in ultra-high energy heavy ion collisions remains under debate. In a previous work~\cite{Oliinychenko:2018ugs} we suggest a microscopic explanation of the deuteron production in central ultra-relativistic Pb+Pb collisions, the main mechanism being $\pi pn \leftrightarrow \pi d$ reactions in the hadronic phase of the collision. We use a state-of-the-art hybrid approach, combining relativistic hydrodynamics for the hot and dense stage and hadronic transport for a later, more dilute stage. Deuteron rescattering in the hadronic stage is implemented explicitly, using its experimentally measured vacuum cross-sections. In these proceedings we extend our previous work to non-central collisions, keeping exactly the same methodology and parameters. We find that our approach leads to a good description of the measured deuteron transverse momentum spectra at centralities up to 40\%, and underestimates the amount of deuterons at low transverse momentum at higher centralities. Nevertheless, the coalescence parameter $B_2$, measured by ALICE collaboration, is reproduced well in our approach even for peripheral collisions.}


\begin{document}

\section{Introduction and Methodology}

\begin{figure}[htb]
\centering
\includegraphics[width=0.55\textwidth]{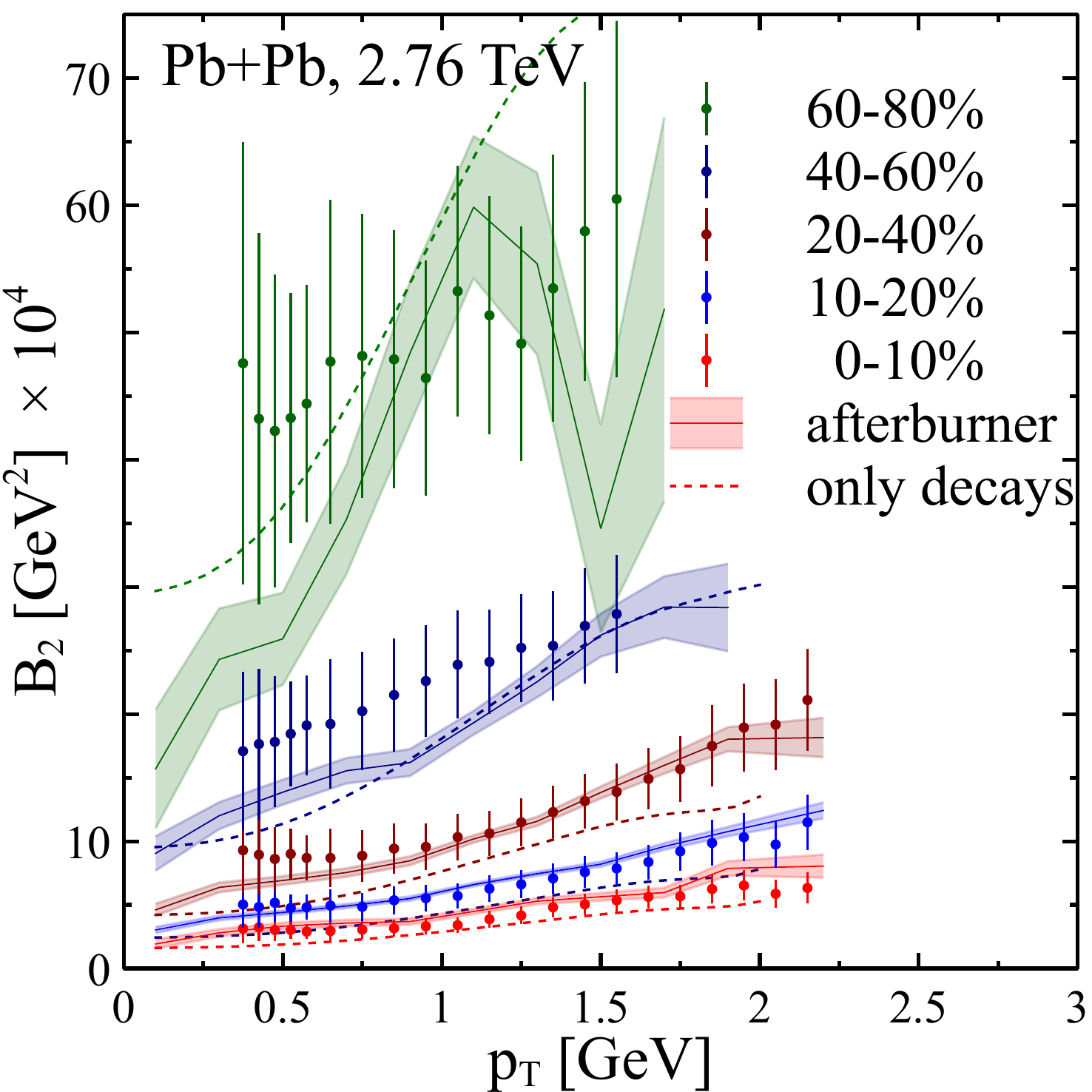}
\caption{Coalescence parameter $B_2$, measured by ALICE collaboration in Pb+Pb collisions at $\sqrt{s_{NN}} = $ 2.76 TeV \cite{Adam:2015vda} (circles) is compared to the outcome of our hybrid (hydrodynamics + transport) approach calculation (lines) and to the same calculation without rescatterings, only with decays (dashed lines). Note that we do not apply coalescence model. Instead, we adopt the approach of~\cite{Oliinychenko:2018ugs}, where deuterons are produced at particlization, similarly to hadrons, and rescatter in the hadronic stage. $B_2$ emerges automatically in this case from dividing deuteron over protons spectra, according to the Eq. \ref{Eq:I}.}
\label{Fig:B2}
\end{figure}

One of the puzzling phenomena observed in the ultrarelativistic ion collisions is the production of the light nuclei out of a hot fireball, with typical temperature much larger than the binding energies of the nuclei. The light nuclei yields and spectra, produced in Pb+Pb collisions at $\sqrt{s_{NN}} = $2.76 TeV per nucleon pair, are measured by ALICE collaboration \cite{Adam:2015vda}. The yields of nuclei at midrapidity $\frac{dN_d}{dy}|_{y=0}$, from deuteron to $^4\mathrm{He}$, as well as anti-nuclei, and a hypernucleus $^3_{\Lambda}\mathrm{H}$ are described well by a thermal model \cite{Andronic:2010qu} via $\frac{dN_d}{dy}|_{y=0} = \frac{g V}{2\pi^2 \hbar^3} T \, m^2 \, K_2\left(\frac{m}{T}\right) \approx \frac{gV}{(2\pi\hbar)^3} (2\pi m T)^{3/2}e^{-m/T}$, where $g$ is  degeneracy, $V$ is the volume of the fireball at the hadronic chemical freeze-out, $m$ is the mass of a nucleus, and $T = 155$~MeV is the chemical freeze-out temperature. The transverse momentum spectra of $\pi$, $K$, $p$, $d$, and $^3\mathrm{He}$ in central Pb+Pb collisions are described by a blast-wave model fit with a kinetic freeze-out temperature $T_{KFO} = 113$~MeV \cite{Adam:2015vda}. Individual blast-wave fits to deuteron spectra in different centrality classes produce kinetic freeze-out temperatures from 77 to 124~MeV \cite{Adam:2015vda}. In this manuscript we attempt to explain the differences between these temperatures from a microscopic perspective.

An alternative way to understand the light nuclei production in the ultrarelativistic ion collisions is a coalescence model. It postulates, that nuclei are formed at a late stage of the expansion from nucleons that reside close in the phase space. 
The coalescence model predicts momentum spectra of nuclei with number of protons $Z$ and number of neutrons $A - Z$ being proportional to the powers of the proton and neutron spectra $\left(E_p \frac{dN_p}{d^3p}\right)^Z \left(E_n \frac{dN_n}{d^3p}\right)^{A-Z}$. Particularly, for deuterons (taking into account that at 2.76 TeV proton and neutron spectra should be identical), after integration over the polar angles:

\begin{eqnarray}
 \frac{1}{2\pi} \frac{d^2 N_d}{p_T dp_T dy} |_{p_T^d = 2 p_T^p} = B_2 (p_T) \left(\frac{1}{2\pi} \frac{d^2 N_p}{p_T dp_T dy} \right)^2 \,,
 \label{Eq:I}
\end{eqnarray}

where $B_2(p_T)$ can in principle be computed ab initio in an elaborate version of a coalescence model~\cite{Scheibl:1998tk}. Here we reproduce $B_2(p_T)$ in a different way.

Our approach has been described in detail elsewhere \cite{Oliinychenko:2018ugs}, and here we only briefly summarize it. We simulate heavy ion collisions using a hybrid approach, combining relativistic hydrodynamics for the denser stage of the fireball expansion, and hadronic transport approach (also called afterburner) for the later, more dilute stage. The initial entropy density distributions in Pb+Pb collisions are given by the Trento Monte Carlo model \cite{Bernhard:2016tnd} with default parameters to approximate the IP-Glasma initial condition \cite{Schenke:2012wb}. The hydrodynamic evolution starts at $\tau_0=0.3$ fm/c with a shear viscosity over entropy density ratio $\eta/s=0.16$. The CLVisc code \cite{Pang:2018zzo} is used to solve the viscous hydrodynamics equations. The s95p-pce lattice equation-of-state \cite{Borsanyi:2012ve} is used, which matches a chemically equilibrated hadron gas at temperatures between 150 and 184 MeV. On the constant temperature hyper-surface with $T_{\rm frz}=155$ MeV, which is equal to the chemical freeze-out temperature in the thermal model, we sample hadrons as well as deuterons using the Cooper-Frye formula \cite{Cooper:1974mv}. These sampled hadrons and deuterons are allowed to rescatter with their vacuum hadronic cross-sections using the SMASH hadronic transport approach \cite{Weil:2016zrk}. We treat deuterons in the afterburner as pointlike particles. The implemented reactions are $\pi d \leftrightarrow \pi p n$, $N d \leftrightarrow N n p$, $\bar{N} d \leftrightarrow \bar{N} n p$, elastic $\pi d$, $N d$ and $\bar{N} d$, and all the corresponding ones for anti-deuterons. The cross-sections can be found in \cite{Oliinychenko:2018ugs}. Our implementation strictly obeys the detailed balance principle. The most important reaction is $\pi d \leftrightarrow \pi p n$ due to the high abundance of pions and large cross-section. In general, the setup for this calculation is exactly the same as in \cite{Oliinychenko:2018ugs}.

\section{Results}

\begin{figure}[htb]
\centering
\includegraphics[width=0.4\textwidth]{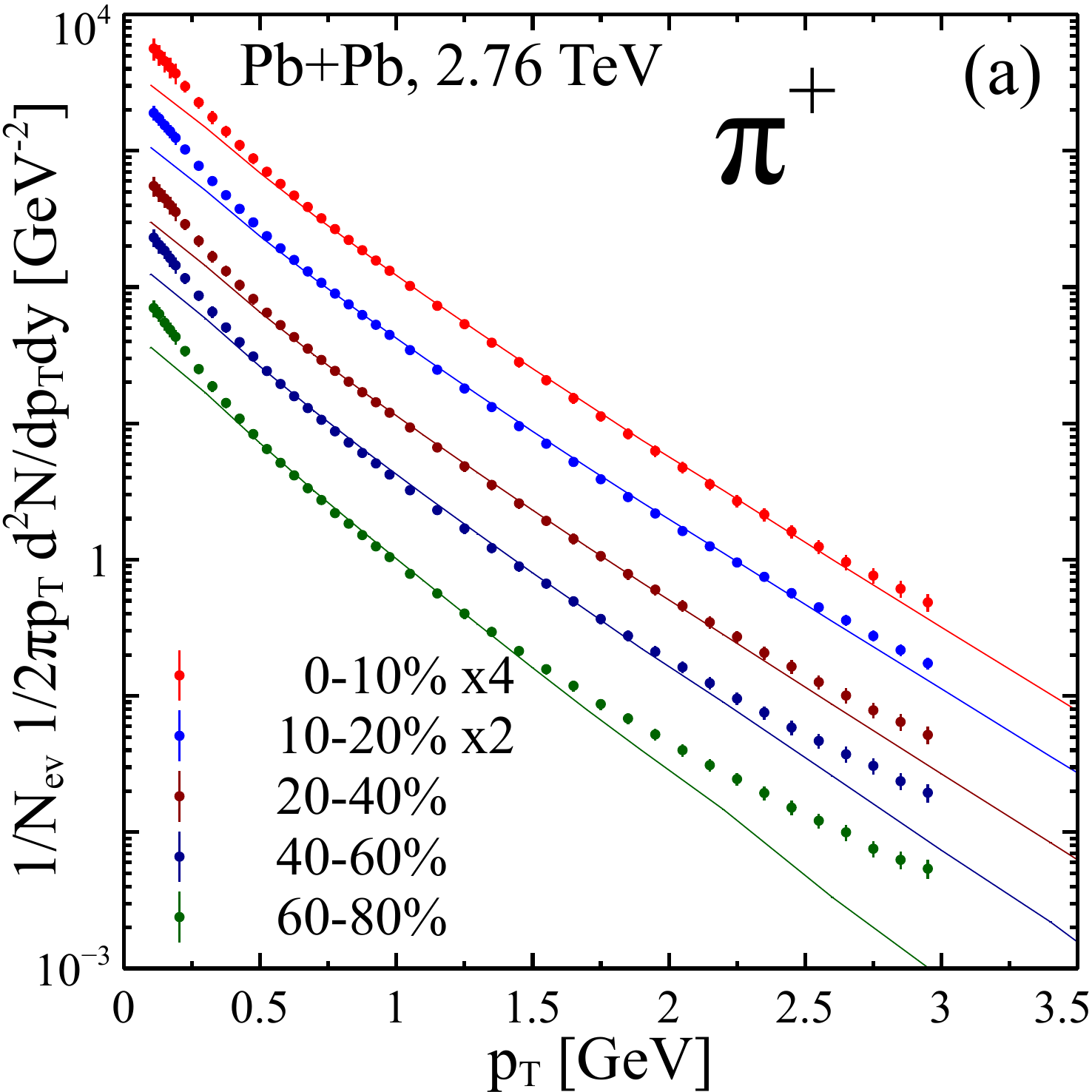}
\includegraphics[width=0.4\textwidth]{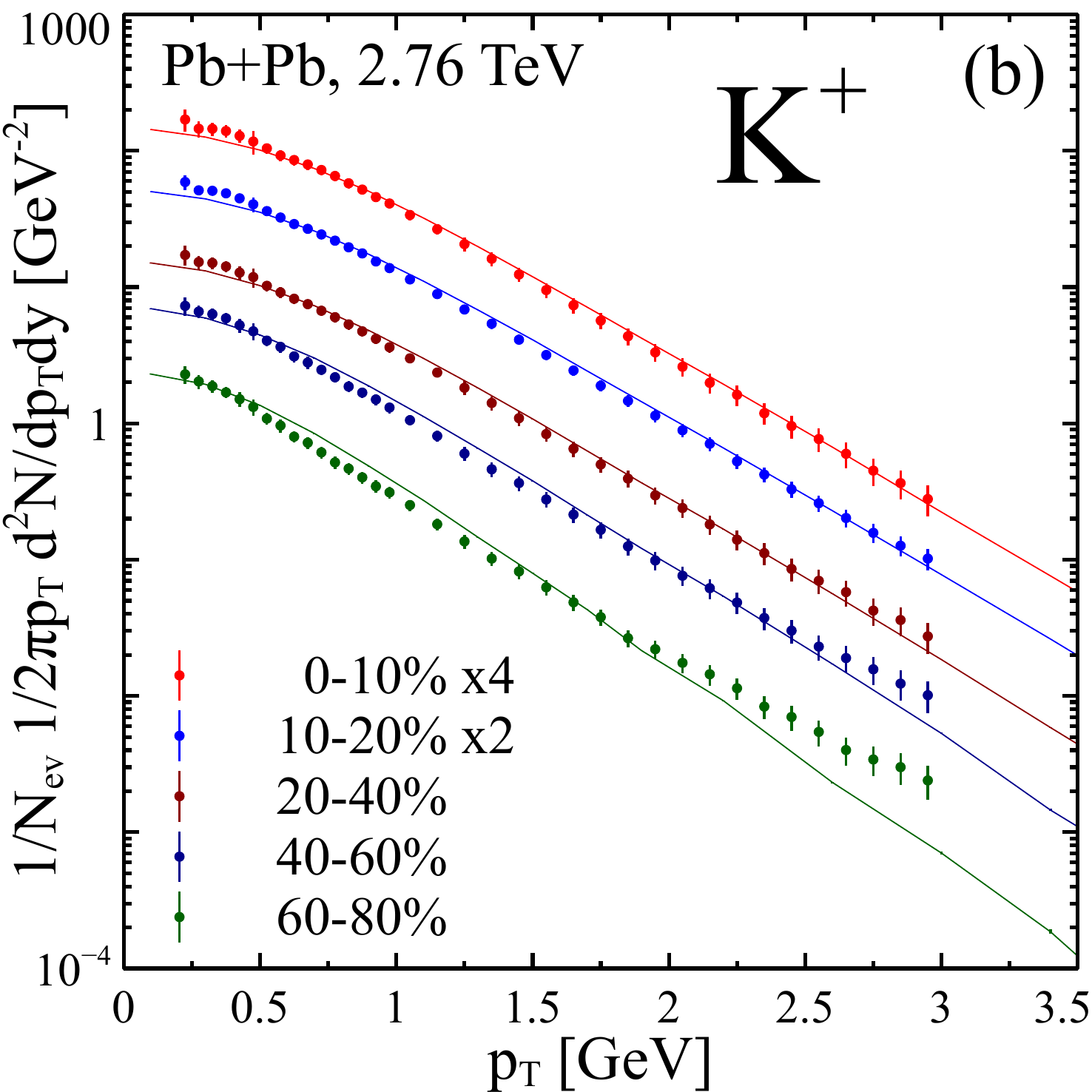}\\
\includegraphics[width=0.4\textwidth]{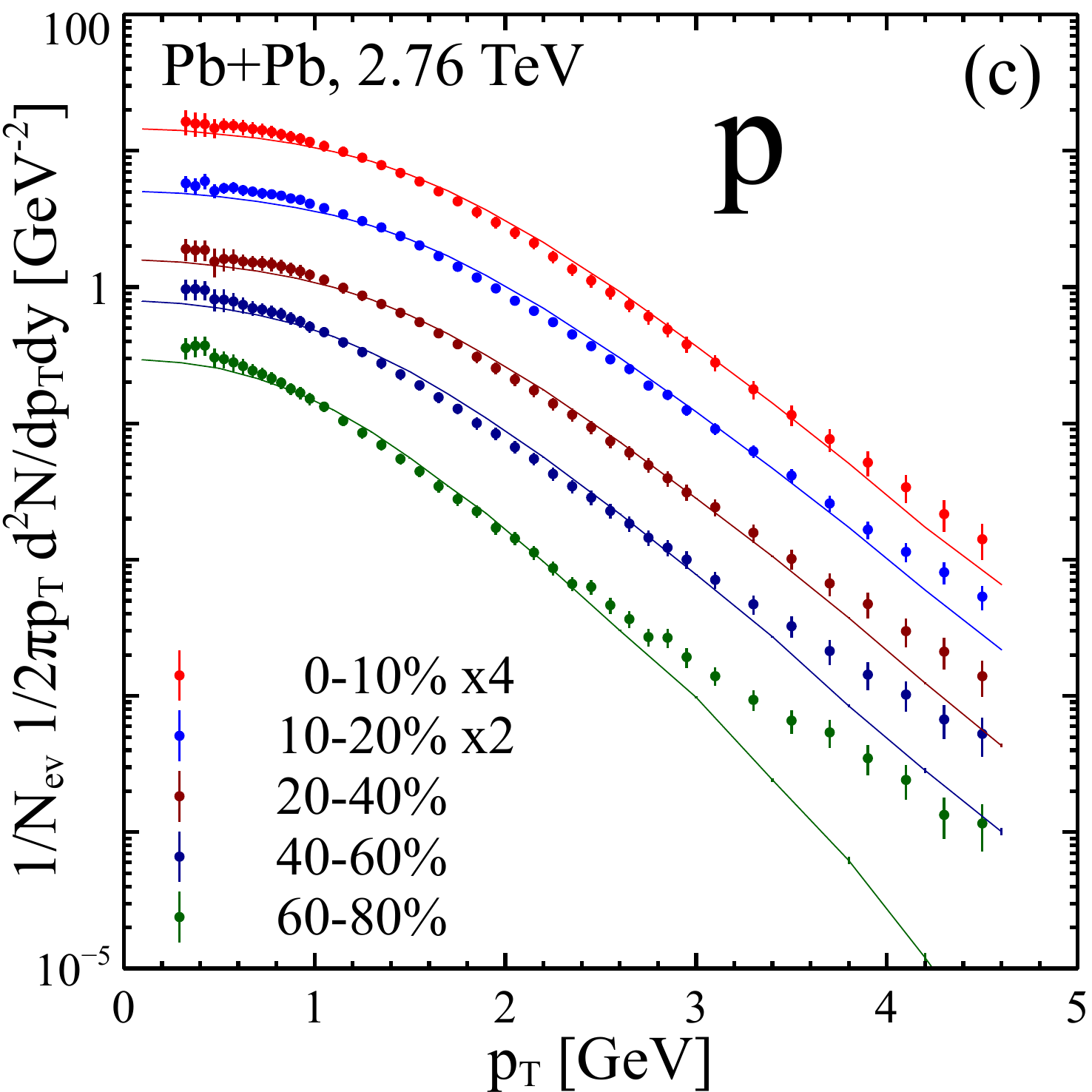}
\includegraphics[width=0.4\textwidth]{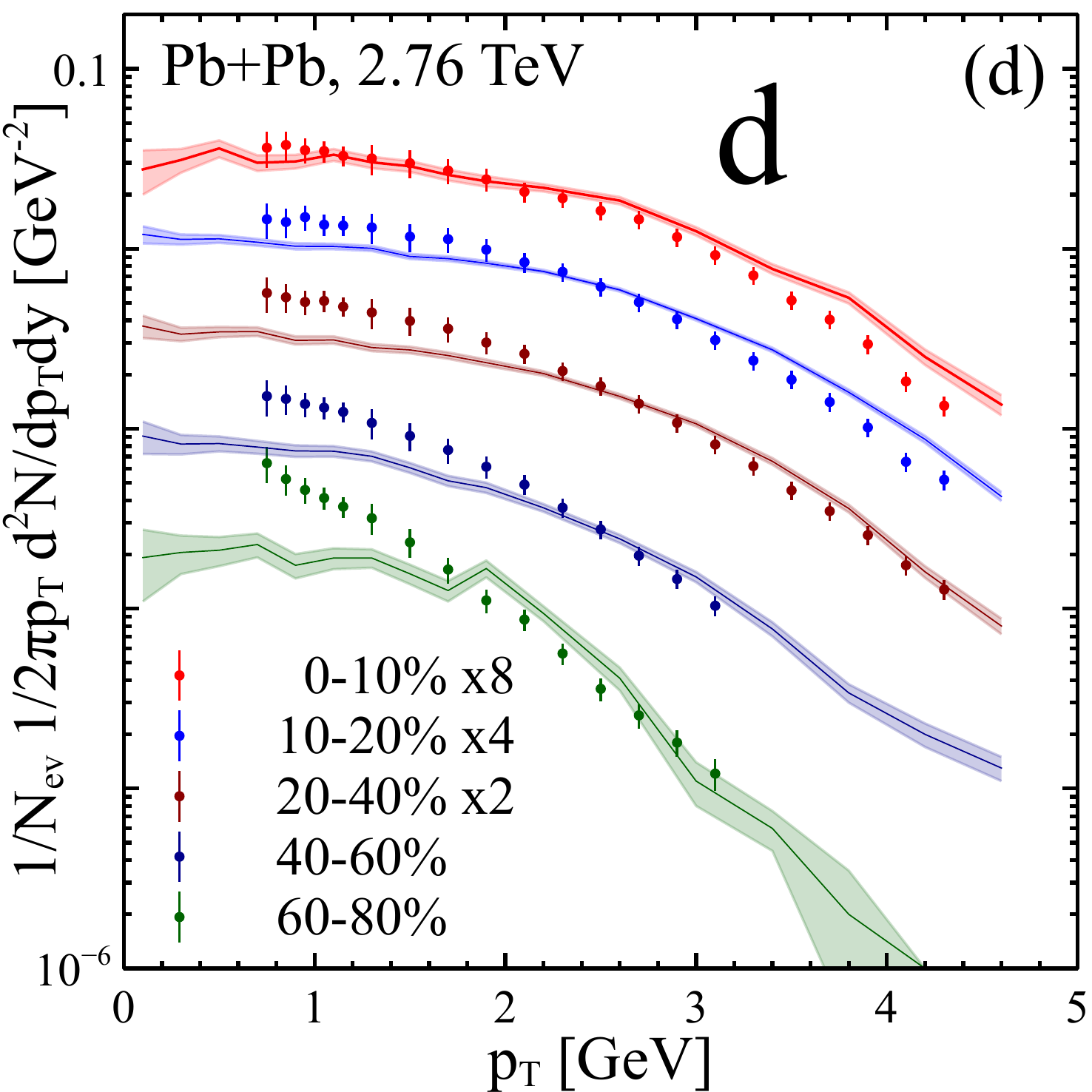}
\caption{Centrality-dependent $p_T$-spectra of identified particles((a) -- pions, (b) -- kaons, (c) -- protons, (d) -- deuterons) in Pb+Pb collisions at $\sqrt{s_{NN}} = $ 2.76 TeV. Measurements by ALICE collaboration \cite{Abelev:2013vea,Adam:2015vda} (circles) are compared to our hybrid (hydrodynamics + transport) approach calculation (lines). Note that computing deuteron spectra is parameter-free and does not apply coalescence model (see text).}
\label{Fig:Spectra}
\end{figure}

The main result of this manuscript is reproducing the experimentally measured behaviour of the $B_2(p_T)$ coalescence parameter of deuterons against collision centrality, as shown in Fig. \ref{Fig:B2}. Our approach correctly captures the growth of $B_2$ against $p_T$ for all centralities, except the most peripheral 60-80\% bin, where the applicability of the hybrid approach in general is doubtful. Nevertheless, even for this most peripheral bin our result is consistent with the data. The $B_2$ itself is a ratio of the $p_T$ spectra, which tends to partly cancel systematic uncertainties, both in experiment and in our simulation. To understand these uncertainties better, it is instructive to analyze and compare the individual $p_T$ spectra too. The $p_T$ spectra of pions, kaons, protons, and deuterons are shown in Fig. \ref{Fig:Spectra}. One can see, that we capture the deuteron spectra at $p_T > 1.5$ GeV rather well for all centralities. At the same time, we underestimate the amount of deuterons at low $p_T$, especially in peripheral collisions. Still, this does not impact the $B_2$ description much, because already $B_2$(0.5 GeV) is connected by Eq. \ref{Eq:I} to the deuteron spectrum at $p_T = 1$ GeV, where the discrepancy is not as large. The underestimation of the deuteron spectra at $p_T < 1.5$ GeV may be caused by the underestimation of the proton spectra at  $p_T < 0.75$ GeV, which can be observed in Fig. \ref{Fig:Spectra}. The pion and kaon spectra, shown in Fig. \ref{Fig:Spectra}, are almost unchanged by the hadronic afterburner, unlike proton and deuteron spectra. Therefore, a fair reproduction of pion and kaon spectra, as in Fig. \ref{Fig:Spectra}, shows that the parameters of the hydrodynamical evolution were chosen reasonably.

In this manuscript we limited ourselves to computing the deuteron transverse momentum spectra in Pb+Pb collisions at 2.76 TeV. It is also interesting to compute flow observables in our approach, which is left for future work. Applying our approach at lower energies and for smaller colliding systems will help to explore the connection between spatial nucleon density fluctuations and light nuclei production, and may eventually provide the information about the critical point of the deconfinement transition \cite{Sun:2018jhg}.

\vspace{6pt}


\funding{ D. O. and V. K. were supported by the U.S. Department of Energy, Office of Science, Office of Nuclear Physics, under contract number DE-AC02-05CH11231 and received support within the framework of the
Beam Energy Scan Theory (BEST) Topical Collaboration.
L.-G.P. was supported by the National Science Foundation (NSF) within the framework of the JETSCAPE collaboration, under grant number ACI-1550228. H. E. acknowledges funding of a Helmholtz Young Investigator Group VH-NG-822 from the Helmholtz Association and GSI.}

\acknowledgments{D. O. thanks I. Karpenko for explaining the effects of hybrid approach parameters on the final hadron spectra, and thanks M. Ploskon for pointing to the ALICE data. We also wish to thank Ulrich Heinz, and Jürgen Schukraft, for their useful and insightful comments. Computational resources have been provided by the Center for Scientific Computing (CSC) at Goethe-University of Frankfurt.}

\conflictsofinterest{The authors declare no conflict of interest.} 

\reftitle{References}

\end{document}